\documentclass[iop]{emulateapj} 

\usepackage{times}
\usepackage{graphics,epsf}
\usepackage{amsmath}                
\usepackage{amsfonts}               
\usepackage{amssymb}                
\usepackage{epsfig}                 
\usepackage{rotating}
\usepackage{color}
\usepackage{tabularx}
\usepackage{url}                       

\newcommand{\AU}{{~\rm AU}}

\def \AU{~\rm{AU}}

\def \apj{ApJ}
\def \aap{A\&A}

\def \mnras{MNRAS}
\def \apjl{ApJ Lett.}
\def \apjs{ApJ Suppl. Ser.}

\def \na{New A}
\def \memsai{Mem,Soc.Astron.Italiana}

\begin{document}

\title{BINARY SYSTEMS OF CORE COLLAPSE SUPERNOVA POLLUTING A GIANT COMPANION}

\author{Efrat Sabach\altaffilmark{1} and Noam Soker\altaffilmark{1}}
\altaffiltext{1}{Department of Physics, Technion -- Israel Institute of Technology, Haifa 32000,
Israel; efrats@physics.technion.ac.il; soker@physics.technion.ac.il}

\begin{abstract}
We examine binary systems where the more massive star,
the primary, explodes as a core collapse supernova (CCSN) the
secondary star is already a giant that intercepts a large fraction
of the ejecta.
The ejecta might pollute the secondary star with
newly synthesized elements such as calcium.
We use Modules for Experiments in Stellar Astrophysics (MESA) to
calculate the evolution of such SN-polluted giant (SNPG) binaries.
We estimate that on average at any given time tens of SNPGs are
present in the Galaxy, and $\approx 10$ SNPG objects are present in
the Magellanic Clouds.
We speculate that the high calcium abundance of the recently discovered
evolved star HV2112 in the Small Magellanic Cloud might be the result of
an SNPG with a super AGB stellar secondary of a mass $\approx 9 M_\odot$.
This rare SNPG scenario is an alternative explanation to HV2112 being a
Thorne-{\.Z}ytkow object (TZO).
\end{abstract}


\section{INTRODUCTION}
\label{sec:introduction}
In a recent work \cite{Schaffenrothetal2015} suggested that the
extreme runaway star HD~271791 was polluted by gas from a
core-collapse supernova (CCSN).
The observed enrichment indicates that HD~271791 had been
ejected by a SN explosion of a very massive compact primary, probably
a Wolf-Rayet (WR) star.
To avoid engulfment during the giant phase of the CCSN progenitor,
the polluted star cannot be too close to the CCSN.
Hence, to intercept a large fraction of the newly synthesized elements
in the CCSN, the companion should be a giant.
We here study some aspects of the evolution of such binary systems.
We set aside the question whether the CCSN ejecta actually removes
a large part of the giant, and no pollution occurs, as claimed by
\cite{Hiraietal2014}.

Such a pollution might account for the presence of rare stars with
peculiar abundances, e.g., HV2112 \citep{Papishetal2015}.
\cite{Levesque2014} found the evolved star HV2112 in the Small
Magellanic Cloud (SMC) to have peculiar abundances.
They suggested that the star is a red supergiant (RSG) star, and
that the peculiar abundances can be understood if HV2112 is a
Thorne-{\.Z}ytkow object (TZO). TZOs are RSG stars that have a
neutron star (NS) at their center \citep{ThorneZytkow1975,
ThorneZytkow1977}.
The star is powered by accretion on to the NS and/or by nuclear
burning in a region away from the NS.
The most likely formation scenario for a TZO in this case is a NS
that in-spiraled inside the envelope of a RSG star, down to the core.
The NS then destroyed the core and replaced the core as the
central dense object. Part of the destroyed core formed a
temporary accretion disk around the NS.

\cite{Toutetal2014} examined whether HV2112 is a TZO or
perhaps a super asymptotic giant branch (SAGB) star.
SAGB stars are stars with a typical initial mass range of
$\approx 7-11M_\odot$ (with dependence on the convective overshoot
treatment; \citealt{Eldridge&Tout2004, Siess2006}) with an oxygen/neon
core undergoing thermal pulses with third dredge up.
\cite{Toutetal2014} argued that SAGBs can synthesize
most of the elements that are used to claim that HV2112 is a TZO through
s-process, e.g., molybdenum, rubidium, and lithium.
However, they found no way for a SAGB star to synthesize calcium.
They suggested that the observed high calcium abundance can be
attributed to its synthesis in the temporary accretion disk around
the NS, composed of the destroyed core material in the TZO
formation process.
In such an accretion disk temperatures and densities are high enough
for calcium nucleosynthesis \citep{Metzger2012}.
The kinetic energy of the disk wind that is required to spread
calcium in the giant has enough energy to unbind the envelope, and
thus \cite{Toutetal2014} postulated that the outflow is collimated,
hence most of it escapes from the star.

However, it is not clear that a TZO can form at all.
Based on earlier studies of common envelope (CE) ejection by jets
\citep{Armitage2000, Soker2004, Chevalier2012}, \cite{Papishetal2015}
studied the removal of the CE that supposedly leads to the
formation of TZOs. \cite{Papishetal2015} found that the jets are
launched by the accretion disk while the NS is still in a Keplerian
orbit around the central part of the core that is still intact.
Therefore, they argued, the jets are not well collimated,
and the envelope and a large part of the core will be ejected.
\cite{Papishetal2015} speculated that the calcium in HV2112 comes
from an explosion of a supernova (SN) while HV2112 was already a
giant star, hence could intercept a large fraction of the SN
ejecta. The exploding star was just slightly more massive than
HV2112 when they both were on the main sequence. In such massive
binary systems the lighter star expands to become a giant before
the more massive star explodes.

In the present paper we examine in more detail the scenario
proposed by \cite{Papishetal2015}.
In this SN-polluted giant (SNPG) scenario we specifically study
binary systems of two stars
that are massive, $\approx 8.5-20M_\odot$, and are very close in
initial masses, $M_{1,0}-M_{2,0} \approx 0.5-1 M_\odot$.
The SN explosion of the primary star might pollute the secondary
if the two stars are not too far apart.
The secondary stars on the more massive end of this range will
result in RSG stars with enhanced newly synthesized elements,
whereas those at the lower end would become polluted SAGBs.
Therefore, we also examine a sub-group of binaries where the
secondary also qualifies as a SAGB star, $M_{2,0}\approx 8.5-11M_\odot$,
and examine the ejecta fraction that might be intercepted by the
secondary star.

In section \ref{sec:evolution} we study the binary evolution of
such systems. In section \ref{sec:HV2112} we discuss a sub-group
of such systems and propose a possible explanation for the calcium
enrichment in HV2112. In section \ref{sec:BR} we estimate the
birthrate of such systems. Our summary is in section
\ref{sec:Summary}.

%
\section{BINARY EVOLUTION}
\label{sec:evolution}

\begin{figure}
\begin{center}
\includegraphics[width=80mm]{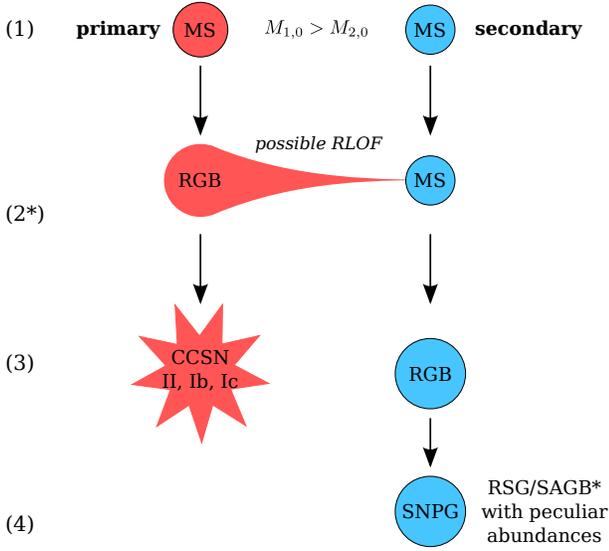}
\caption{
Schematic evolution of the SN-polluted giant (SNPG) binary systems studied
here, where the initial mass of the primary (red), $M_{1,0}$, is slightly
larger than that of the secondary (blue), $M_{2,0}$.
The evolution is plotted from ZAMS (row 1)
until the primary star explodes as a core collapse supernova (CCSN)
(row 3).
During the explosion the primary chemically pollutes the red super giant
(RSG) secondary star which then becomes a SN-polluted giant (SNPG) (row 4).
Roche lobe overflow (RLOF) is possible
once the primary is a red giant branch (RGB) star (row 2).
This is possible for an initial separation of $a_0 \la 2.5 R_{1, {\rm RGB}}$,
where $r_{1,{\rm RGB}}$ is the maximum radius of the primary during its
RGB phase.
In such cases the post-transfer secondary star , $M_{2,PT}$, is
the more massive star in the system since the two stars are of close
initial masses (row 3).
Since this possible early mass transfer episode does not influence much
the final SNPG outcome, we do not study it here in much detail.
The resulting secondary star will be a super asymptotic giant branch
(SAGB) star in case its initial mass is $\approx 8.5-11M_\odot$.
}
\label{fig:mech}
\end{center}
\end{figure}

We examine the evolution of massive binary systems where on the
zero age main sequence (ZAMS) the primary of mass $M_{1,0}$ is
slightly more massive than the secondary of mass $M_{2,0}$, as
schematically presented in the first row of Fig. \ref{fig:mech}.
We are interested in primary stars that end as CCSNe, which implies
an initial primary mass of $M_{1,0} \ga 9 M_\odot$.
To intercept a large fraction of the newly synthesized elements in the CCSN
of the progenitor, the companion should be a an evolved giant while
the explosion of the primary takes place (third row in Fig. \ref{fig:mech}).
The ejecta from the exploding primary star pollutes the secondary
which becomes a SNPG (fourth row in Fig. \ref{fig:mech}), as has been
suggested for example for the hyper--runaway star HD271791
\citep{Schaffenrothetal2015}.
We found that for primary stars in the mass range of
$M_{1,0}\approx 9-20M_\odot$ the secondary must be in the mass range
of $M_{1,0}>M_{2,0}\ga M_{1,0} - 1M_\odot$, to allow for the SNPG
scenario (more details in section \ref{sec:BR}).

The initial orbital separation cannot be too small since we must
avoid the possibility that the secondary would be engulfed by the primary
during the primary giant phase.
Yet we point out that for the case of close binaries it  is possible for
the primary to fill its Roche Lobe during its red giant branch (RGB)
phase and transfer some of its outer envelope to its less massive companion
(second row of Fig .\ref{fig:mech}).
We find that for our close systems
$\left( q\equiv M_{2}/M_{1}\approx 1\right)$ RLOF occurs for 
$R_{1,\rm RGB}/a_0\lesssim 0.4$ \citep{Eggleton1983},
where $R_{1,\rm RGB}$ is the maximum radius of the primary on the RGB.
Hence,  assuming the initial orbital separation of the system is
$a_0\la 2.5 R_{1,\rm RGB} $,
the primary might fill its Roche lobe and transfer the outer layers of
its envelope to the secondary star, while the latter is still on the MS.
Due to the close initial mass of the stars, in case of mass
transfer the post-transfer primary continues to evolve with a lower
mass envelope, and the post-transfer (PT) secondary evolves as a star slightly
more massive than the initial primary, $M_{1,PT}<M_{2,PT}$.
Moreover, since the stars are of close initial masses, as the
the secondary grows to be the more massive star in the system the
orbital separation grows and the mass transfer might cease,
at least for some period.
We do not go into details of such an early-RLOF phase since it has little
importance to the pollution of the secondary and the SNPG outcome.
In addition, it is possible that later the secondary will fill its Roche lobe
before the primary explodes. If this occurs we might form a common
envelope. This is not studied here.

The evolutionary scheme presented in Fig. \ref{fig:mech} is
significantly different from those in \cite{SabachSoker2014} as
here there is no reverse evolution, the primary experiences a CCSN
explosion, and the secondary is a giant when the primary ends its
evolution; these don't hold in the scenarios discussed in
\cite{SabachSoker2014}.

To follow the evolution of each star from zero age main sequence (ZAMS)
we use the Modules for Experiments in Stellar Astrophysics (MESA),
version 7184 \citep{Paxton2011}, for non-rotating stars.
For the lower mass range of the systems studied here we were
able to evolve the stars until the formation of an ONe core. 
We encountered some numerical difficulties at very late evolutionary
stages because off-center burning flames make the computation numerically
expensive.
The omission of the final core collapse has no consequences for our
study.

\begin{figure}
\begin{center}
\includegraphics[width=85mm]{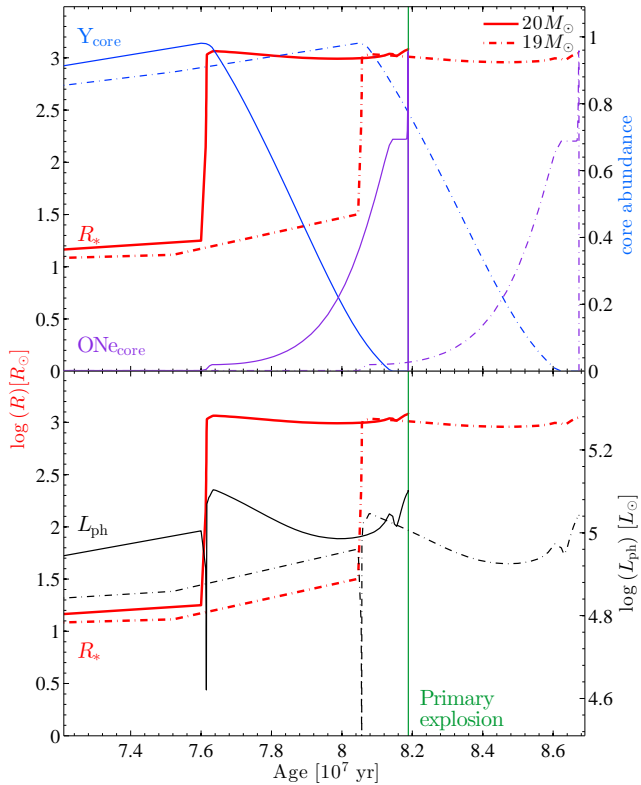}
\caption{The evolution of a representative system
evolved using MESA from ZAMS with solar metallicity ($z=0.02$).
The primary initial mass is $M_{1,0}=20M_\odot$ and the secondary
initial mass is $M_{2,0}=19M_\odot$.
Each star was evolved until explosion as a core collapse supernova.
We present only the late stages of evolution.
The solid and dashed lines represent the primary and secondary stars,
respectively.
Upper panel: The thick-red, blue and purple lines represent the
stellar radius (left axes), He fraction in the core, and ONe fraction in
the core (right axes) ,respectively.
Lower panel: The thick-red and black lines represent the stellar radius
(left axes), and stellar luminosity (right axes), respectively.
The primary explosion is marked with a green vertical
line, corresponding to the third row in Fig. \ref{fig:mech}.
The primary reaches its final stages of evolution when the secondary is
a red giant star.
At the late evolutionary stages of $M_2$ it becomes polluted red duper giant
(RSG) with peculiar abundances (forth row in Fig. \ref{fig:mech}).
}
\label{fig:binary_ev}
\end{center}
\end{figure}

Fig \ref{fig:binary_ev} shows the evolution of a representative binary
system of the higher mass range evolving according to our assumptions.
The initial primary and secondary masses are $M_{1,0}=20M_\odot$
and $M_{2,0}=19M_\odot$, respectively.
The system was evolved from ZAMS with solar metallicility ($z=0.02$),
and until the explosion of each star.
It is evident that once the primary explodes the secondary is
already a red-giant.
It is also apparent that during the giant phase of the primary star
mass transfer is possible via RLOF, depending on the initial separation
of the system (see above).
This possibility is not presented here as there is little significance
on the SNPG outcome.

We note that there are uncertainties as to whether SN ejecta can
enrich a giant companion  star.
\cite{Hiraietal2014} find in recent numerical simulations that
the shock propagating through the secondary by the SN ejecta
can heat the companion.
This might lead to the removal of up to $25 \%$ of the companion mass
by the excess energy in case of a close binary, and could rule out the
proposed model.
This difficulty might be overcome by non-spherical SN ejecta with a
large concentration of calcium and other synthesized elements ejected
towards the companion.
Another process that can overcome the difficulties posed by the results
of \cite{Hiraietal2014} and allow large quantities of calcium and other
heavy elements to be accreted on to the companion is if the newly
synthesized elements from the core of the SN expand in dense clumps.
Such clumps can penetrate deeper to the star, and stay bound.
An  estimation of the overall ejecta fraction intercepted
by the secondary star is calculate next in section \ref{sec:HV2112} for
the case of a SAGB companion.
\section{HV2112 as a SAGB star}
\label{sec:HV2112}

\begin{figure}
\begin{center}
\includegraphics[width=80mm]{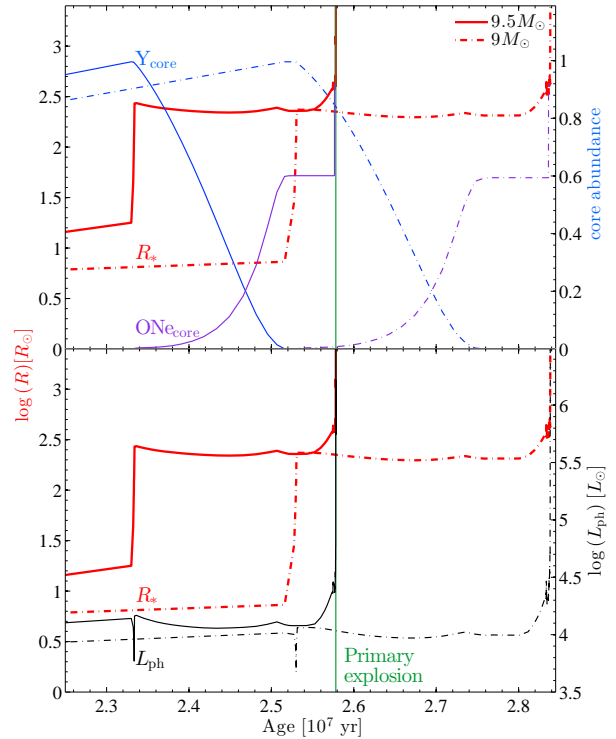}
\caption{  The evolution of a representative system that might
account for HV2112. The primary initial mass is $M_{1,0}=9.5M_\odot$ and
the secondary initial mass is $M_{2,0}=9M_\odot$.
We start the evolution from the ZAMS with SMC metallicity (z=0.004) but
show here only the late stages.
Each star was evolved until late evolutionary stages prior to the explosion
as a core collapse supernova.
The solid and dashed lines represent the primary and secondary stars,
respectively.
Upper panel: The thick-red, blue and purple lines represent the
stellar radius (left axes), He fraction in the core, and ONe fraction in
the core (right axes) ,respectively.
Lower panel: The thick-red and black lines represent the stellar radius
(left axes), and stellar luminosity (right axes), respectively.
The primary explosion is marked with a green vertical
line, corresponding to the third row in Fig. \ref{fig:mech}.
The primary reaches its final stages of evolution when the secondary is
a red giant star.
At the late evolutionary stages of $M_2$ it becomes a super asymptotic giant
branch (SAGB) star with an ONe core (forth row in Fig. \ref{fig:mech}).
}
\label{fig:binary_HV2112}
\end{center}
\end{figure}

We examine a sub-group of SN-polluted giants (SNPGs) where the mass
of the secondary during the explosion of the primary is in the range
of $\approx 8.5-11M_\odot$, in order to qualify as a SAGB star
\citep{Eldridge&Tout2004, Siess2006}.
We present here a representative case for such systems that might account
for HV2112 being a SAGB star.
For the SMC metallicity ($Z\sim0.004$; \citealt{Diagoetal2008})
\cite{Dohertyetal2015} find that SAGB stars are in an initial mass range
of  $7.1-8.8M_\odot$.
Accordingly we chose the initial primary and secondary masses, as shown in
Fig .\ref{fig:binary_HV2112}, to be
$M_{1,0}=9.5M_\odot$ and $M_{2,0}=9M_\odot$, respectively.
We evolved each star with an SMC metallicity of $z=0.004$ using MESA.

The initial stellar masses were chosen according to four criteria:
(1) The primary is massive enough to explode as a CCSN
(triggered by electron capture).
(2) The stars must be of close initial masses for the secondary to be an
RGB star during the explosion of the primary.
(3) The secondary must qualify as a SAGB star at late stages.
(4) The evolved SAGB secondary must agree with the properties of HV2112,
e.g., luminosity of $\approx4.6\times 10^4 - 1.1\times10^5 L_\odot$
\citep{Levesque2014, Toutetal2014}.

We note the that during the final (few $\times10^3$ yr) stages of evolution the
stars seem to exceed the Eddington luminosity calculated for electron
scattering opacity.
One should take into account that during this stage the opacity in the
photosphere is much lower than that for electron scattering. Namely, the
star does not reach the Eddington luminosity at the photosphere.
Above the photosphere electron scattering might dominate, and mass loss rate must
be very high.
Our simulations do not include such an enhanced mass loss rate, but this
is what we expect to occur.

As the primary star explodes it chemically pollutes the secondary star,
now an RGB star with a radius of $\approx 230R_\odot$.
The polluted secondary continues to evolve into a SAGB star.
To account for the calcium abundance in HV2112 \cite{Papishetal2015}
assumed that the giant secondary star intercepted a large enough fraction
of the ejecta. We here demonstrate that this is possible.
\cite{Toutetal2014} estimate the cailcium mass in HV2112 to be
$\approx \rm 10^{-4}M_\odot$ from the line ratios presented by
\cite{Levesque2014} for the SMC metallicity.
For massive stars, $9M_\odot\lesssim M_1\lesssim 20M_\odot$, with
solar metallicity exploding as a CCSN the ejected $\rm ^{40}Ca$ mass
is $\simeq\rm few\times10^{-3}-\rm few\times10^{-2}M_\odot$.
This is up to one hundred times the calcium mass estimated in HV2112.
We note that the Ca abundance in the lower range of massive stars
$8M_\odot\lesssim M\lesssim11M_\odot$, exploding as
CCSN hasn't been studied thoroughly for the metallicity of the SMC,
hence we use the above estimations of the ejecta in our calculations.

Overall a fraction of $f\approx 0.01$ of the SN ejecta
must be accreted on to the secondary.
For an accretion efficiency $\eta$ we have
\begin{equation}
f=\frac{1}{4}\left( \frac{R_2}{a} \right)^2\eta=0.01
\label{eq:fraction}
\end{equation}
where $a$ and $R_2$ are the separation of the system and the secondary
radius during the SN explosion, respectively.
To account for $\simeq10^{-4}M_\odot$ of
$\rm ^{40}Ca$ in HV2112 requires that  $\eta\approx0.25$.
In cases where RLOF is avoided at an earlier stage (see section \ref{sec:evolution})
$R_2/a\gtrsim0.4$ hence  $\eta\gtrsim 0.25$.

\section{BIRTHRATE ESTIMATION}
\label{sec:BR}

To estimate the Galactic and Magellanic Clouds (MCs) birthrate of the
studied SNPG systems, and the fraction to all potential progenitor of
CCSNe we proceed as done by \cite{SabachSoker2014}.
For the initial mass function of the relevant primary we take
\citep{Kroupa1993}
\begin{equation}
\frac {d N}{d M} = A M^{-2.7}, \quad {\rm for} \quad 1.0M_\odot < M ,
\label{eq:IMF1}
\end{equation}
where A is a constant.
For the systems studied here we demand that the initial stellar mass must
be $$M_{1,0} > M_{2,0} \gtrsim M_{1,0} - \Delta M \equiv M_{2,0,{\rm min}}.$$
We also assume that the secondary mass distribution is constant in the
allowed range
\\$dN_2=dM_2/M_1$ for $M_1>M_2>0$.
The number of relevant binary systems is given by
\begin{equation}
N_{b} \simeq \int   \frac {d N_1}{d M_{1,0}} \left( \frac{M_{1,0}-
M_{2,0,{\rm{min}}}}{M_{1,0}} \right)dM_{1,0}.
\label{eq:Pr1}
\end{equation}
We find that for primary stars with initial mass of
$15M_\odot \gtrsim M_{1,0} \gtrsim 9M_\odot$ the secondary must be
in the mass range of $M_{1,0} > M_{2,0} \gtrsim M_{1,0} - 0.6M_\odot$,
to allow our evolutionary scenario.
For primary stars with initial mass of
$20M_\odot \gtrsim M_{1,0} \gtrsim 15M_\odot$ the secondary must be
in the mass range of $M_{1,0} > M_{2,0} \gtrsim M_{1,0} - 1.2M_\odot$.
This gives
$$
N_{b} \simeq \int^{15M_\odot}_{9M_\odot}   \frac {d N_1}{d M_{1,0}}
\left( \frac{0.6 M_\odot}{M_{1,0}} \right)dM_{1,0}+
$$
\begin{equation}
\int^{20M_\odot}_{15M_\odot}   \frac {d N_1}{d M_{1,0}} \left(
\frac{1.2 M_\odot}{M_{1,0}} \right)dM_{1,0}=6\times10^{-4}AM_\odot^{-1.7}.
\label{eq:Pr2}
\end{equation}
For the progenitors of CCSNe we take all stars with initial mass
$ M \gtrsim 9M_\odot$, for which integration gives
$N_{\rm CCSN} = 0.014AM_\odot^{-1.7}$.

\cite{Raghavanetal2010} estimate a lower limit of $75\%$ for O-type
stars to have companions. We take a typical fraction $f_{b} \simeq 0.8$
of O-type stars to be in binary systems with an orbital separation
less than $700AU$,
and the relevant binary population of massive stars to span over a range of
4 orders of magnitude (from $a_{\rm min} \simeq 0.1$ to $a_{\rm max}
\simeq 1000\AU$) with an equal probability in the logarithmic of the
orbital separation.
For the pollution of the secondary by the primary to take
place, the  relevant orbital separation distribution is
$a\simeq 2-6R_2\simeq 500-1500R_\odot$ (as we can not completely rule out
some mass transfer, accounting for the lower limit).
We find the orbital separation distribution to span over $\simeq 0.5$ dex.

Accordingly, the probability of a binary system to be in the desired
orbital separation is $f_s\simeq0.5/4=1/8$.
This is a crude estimate as we took an order of magnitude value for the
relevant orbital separation range.
We also note that if the eccentricity of the system is considered the
initial separation range is even larger.

The fraction of the SNPG systems studied here to the number of CCSNe
progenitors is
\begin{equation}
\frac{N_{\rm systems}}{N_{\rm CCSN}} \simeq  \frac { N_{b}}{N_{\rm CCSN}}
f_{b} f_s \simeq 4.3\times 10^{-3}.
\label{eq:frac1}
\end{equation}

Using the CCSN rate in the Galaxy $\simeq0.014 \rm{yr}^{-1}$
\citep{Cappellaroetal1997}, we estimate the Galactic birthrate of
such systems to be  $ \simeq 6 \times 10^{-5} \rm{yr}^{-1}$.
From the lifetime of the polluted secondary stars studied here
from the primary explosion to the secondary explosion
, $\simeq 10^6\rm{yr}$ ( Fig \ref{fig:binary_ev}), we estimate that
on average $\approx 60$ such SNPG systems exist in the Galaxy at any given
time.
\cite{Maoz&Badenes2010} estimate the SN (SNIa+CCSN) rate in the MCs to
be $2.5-4.6\times10^{-3}{\rm yr}$, and the CCSN rate to be $\simeq 2.5$
times more than the SNIa rate.
From the lifetime of the polluted secondary star we estimate that on
average $\approx 10$ such systems exist in the MCs at any given time.

For the sub-case of SAGB polluted giants the
recurrence of objects such as HV2112 is smaller since the SAGB lifetime
is a few $10^5\rm{yr}$  (Fig \ref{fig:binary_HV2112}; \citealt{Doherty2014}).
We estimate that only 1-3 objects will be found in the SMC.
If a large number of such objects will be found, probably neither
our model nor the TZO model will be applicable.
We note that the uncertainties in our estimates are very large,
but non-the-less the conclusion that such SNPG objects are rare is robust.

\section{DISCUSSION AND SUMMARY}
\label{sec:Summary}

We have studied some properties of binary systems where the
secondary star is already in its giant phase when the primary
explodes as a core collapse supernova (CCSN; Fig. \ref{fig:mech}).
For this to occur the initial mass of the secondary star should be
only slightly, by about $5 \%$ and less, below the initial mass of
the primary.

From the results of section \ref{sec:BR} this case occurs in about
$2 \%$ of all CCSNe. This implies that observations of the
post-explosion site can reveal that a giant star still exists
there. With better sky coverage and more SNe in relatively close
galaxies, such cases must be eventually detected.

We then discussed a specific type of such systems where the
orbital separation is such that the secondary star can intercept a
large fraction, about $1 \%$, of the SN ejecta. If the secondary
envelope is not completely ablated by the ejecta, the secondary
becomes polluted with metals from the SN; we term this a
SN-polluted giant (SNPG) scenario. In section \ref{sec:BR} we
estimated the SNPG scenario might occur in about $0.4 \%$ of all
CCSNe. The secondary then lives for some time before it explodes.
We estimated that at any given time there are about 60 SNPG stars
in the Galaxy and about 10 SNPG stars in the Magellanic clouds.
If we allow for a lower chemical pollution, then the orbital
separation can be larger and the number of SNPG systems increases.

We used the SNPG scenario to address the large calcium abundance
of the evolved star HV2112 in the Small Magellanic Cloud (SMC).
Tout et al (2014) find that a super-AGB star can account for the
peculiar abundances of HV2112 besides that of calcium.
\cite{Levesque2014} and \cite{Toutetal2014} argued that the high
calcium abundance is best explained if HV2112 is a
Thorne-{\.Z}ytkow object (TZO). \cite{Papishetal2015}, on the
other hand, argued that it is impossible to bring a NS to the
center of a giant star since the entire envelope and part of the
original core will be ejected by the jets that are launched by the
NS as it in-spirals \citep{Armitage2000, Soker2004,
Chevalier2012}. Instead, \cite{Papishetal2015} speculated that the
high calcium abundance might be explained by pollution
(enrichment) from a more massive companion that had already
exploded as a CCSN.
Based on our finding we conclude that the SN pollution scenario is
much more likely than the TZO explanation for HV2112.

The evolutionary routes discussed here and in \cite{Papishetal2015}
have potential relation to some exploding and erupting astrophysical
objects.
\newline
(1) \emph{Explosion with jets.} The process by which the
in-spiraling NS launches jets and ejects the envelope and core is
very rapid, and will be observed as an explosion
\citep{Chevalier2012}. It is quite likely that \emph{all} CCSNe
have exploded by jets launched by the newly formed NS (or black
hole) at their center, as in the jittering-jet mechanism
(\citealt{Papish&Soker2014}; \citealt{GilkisSoker2014}).
Hence, the NS-core merger might be
wrongly attributed to a CCSN with large pre-explosion mass loss
\citep{Chevalier2012}.
\newline
(2) \emph{Intermediate Luminosity Optical Transient (ILOT) 1.}
If the NS ejects the entire envelope but does not merge with the
core of the RSG star, then the outburst will be much less energetic.
The total energy will be of the order of the binding energy of the
envelope. The luminosity of the outburst will be much below that of
a SN, but more than that of a nova.
The outburst will be classified as an ILOT event. We note that the
in-spiraling of a WD companion to the core of a giant star can also
lead to an ILOT event \citep{SabachSoker2014}.
\cite{Tylendaetal2013} already suggested that the transient
OGLE-2002-BLG-360 is an ILOT (which they termed a red transient)
in the process of a final merger of a common envelope evolution.
\newline
(3) \emph{ILOT 2.}
If an early stage of RLOF takes place once the primary is a red giant,
the mass transfer from the more evolved star to the secondary
star occurs in an unstable manner over a short time, tens of years and
less, this event can be classified as a luminous blue variable
(LBV) major event, or as an ILOT \citep{KashiSoker2010}.
\newline
(4) \emph{Peculiar Superluminous CCSN.} In the proposed scenario
for SNPG the secondary is already a giant when the primary
explodes. It is quite possible that the secondary will be as
bright as the primary. For example, in the binary system
presented in Fig. \ref{fig:binary_ev} the luminosities of the two
stars just before the primary explosion are
$L_1=1.26\times10^5L_\odot$ and $L_2=1.02\times10^5L_\odot$.
We expect also massive CSM, as not all mass is accreted by the
secondary. The collision of SN ejecta with the CSM will form a
superluminous SN, but the presence of a post-explosion giant
remnant will make it a peculiar CCSN.

Our present study adds to the variety of peculiar astrophysical
objects that might be related to peculiar eruptions and explosions.

\bigskip

\textit{Acknowledgements.}
We thank J. J. Eldridge for very helpful
and detailed comments that substantially improved the manuscript.
This research was supported by the Asher Fund for Space Research
at the Technion, and the US-Israel Binational Science Foundation.

\label{lastpage}

\end{document}